\documentclass[twocolumn,twoside,slac_two]{revtex4}
\usepackage{graphicx}
\usepackage{fancyhdr}
\usepackage{url}
\usepackage{booktabs}

\begin{document}

\title{Six-Dimensional Cooling Simulations for the Muon Collider}

%

\author{Pavel Snopok}
\author{Gail Hanson}
\affiliation{Department of Physics and Astronomy, University of California Riverside, Riverside, CA 92521, USA}

\begin{abstract}
The two cooling channels based on the RFOFO ring concept are considered and simulated. One of them is the RFOFO helix, also known as the Guggenheim. The helical shape of the channel resolves the injection and extraction issues as well as the absorber overheating issue. The issue of the RF breakdown in the magnetic field is addressed in the so-called open cavity cooling channel lattice with magnetic coils in the irises of the RF cavities. The details of the tracking studies of both channels are presented and compared to the performance of the original RFOFO cooling ring design.
\end{abstract}

\maketitle

\thispagestyle{fancy}


\section{RFOFO Cooling Ring}

In a Muon Collider design the muon beam 6D phase space volume must be reduced several orders of magnitude in order to be able to further accelerate it. Ionization cooling is currently the only feasible option for cooling the beam within the muon lifetime ($\tau_0 = 2.19$ $\mu$s). The RFOFO ring \cite{PhysRevSTAB.8.061003,RFOFO_ring_2003} is one of the feasible options currently under active investigation along with other designs \cite{HCC,RFOFO_snake,Quad_channel}. The RFOFO ring provides a significant reduction in the six-dimensional emittance in a small number of turns with a relatively low particle loss factor. 6D cooling is achieved by employing the concept of emittance exchange. When a dispersive beam passes through a wedge absorber in such a way that higher momentum particles pass through more material, both the longitudinal and the transverse emittances are reduced. However, the design of the injection and extraction channels and kickers is very challenging for the RFOFO, and the ring could not be used as is because the bunch train is too long to fit in the ring. Both problems would be removed in the RFOFO helix, also known as the Guggenheim channel \cite{Guggenheim}. In addition, using the helix solves another important issue, namely, the overheating in the absorbers.

\begin{figure}
	\centering
		\includegraphics[width=0.48\textwidth]{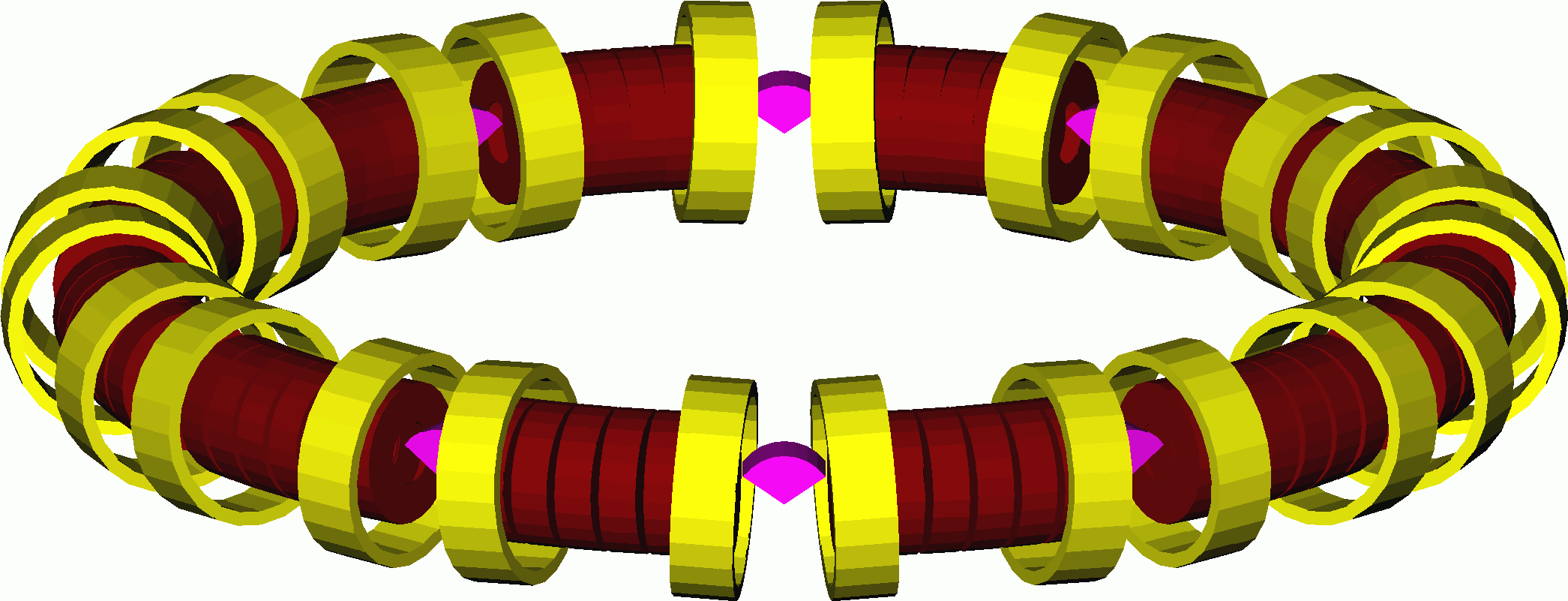}
	\caption{RFOFO ring layout. Yellow---tilted magnetic coils with alternating currents to provide necessary bending and focusing and generate dispersion, magenta---wedge absorbers for cooling and emittance exchange, red---RF cavities for restoring the longitudinal component of the momentum.}
	\label{fig:rfofo}
\end{figure}

The main parameters of the original RFOFO design are summarized in Table~\ref{tab:params} and compared to the parameters of the Guggenheim channel. The layout of the RFOFO ring is shown in Fig.~\ref{fig:rfofo}. The results of particle tracking through the RFOFO channel in the code G4Beamline \cite{g4beamline} are used as the point of reference for comparing the RFOFO and Guggenheim channel efficiencies.

\begin{table}
	\centering
\begin{footnotesize}
		\begin{tabular}{c|c|c|c}
			& & RFOFO	& Guggenheim\\
		  \hline\hline
			RF frequency & [MHz] & 201.25	& 201.25\\
			RF gradient & [MV/m] & 12.835 & 12.621\\
			Maximum axial field & [T]	& 2.77 & 2.80\\
			Pitch & [m] & 0.00 & 3.00\\
			Pitch angle & [deg]	& 0.00 & 5.22\\
			Circumference & [m]	& 33.00	& 32.86\\
			Radius & [m]	& 5.252 & 5.230\\
			Coil tilt (wrt orbit) & [deg]	& 3.04 &3.04\\
			Average momentum & [MeV/c] & 220 & 220\\
			Reference momentum & [MeV/c] & 201 & 201\\
			Absorber angle & [deg] & 110 & 110\\
			Absorber vertical offset & [cm] & 9.5 & 9.5\\
			Absorber axial length & [cm] & 27.13 & 27.13\\
		\end{tabular}
\end{footnotesize}
	\caption{Parameters of the RFOFO ring compared to the Guggenheim helix.}
	\label{tab:params}
\end{table}

\section{Guggenheim Helix}

The layout of the Guggenheim channel to a large extent repeats the one of the RFOFO ring, except for the three meters of separation between the layers of the helix. As a result, the circumference of the helix has to be slightly smaller than that of the ring to keep the arclength of one revolution intact. 

\begin{figure}
	\centering
		\includegraphics[width=0.48\textwidth]{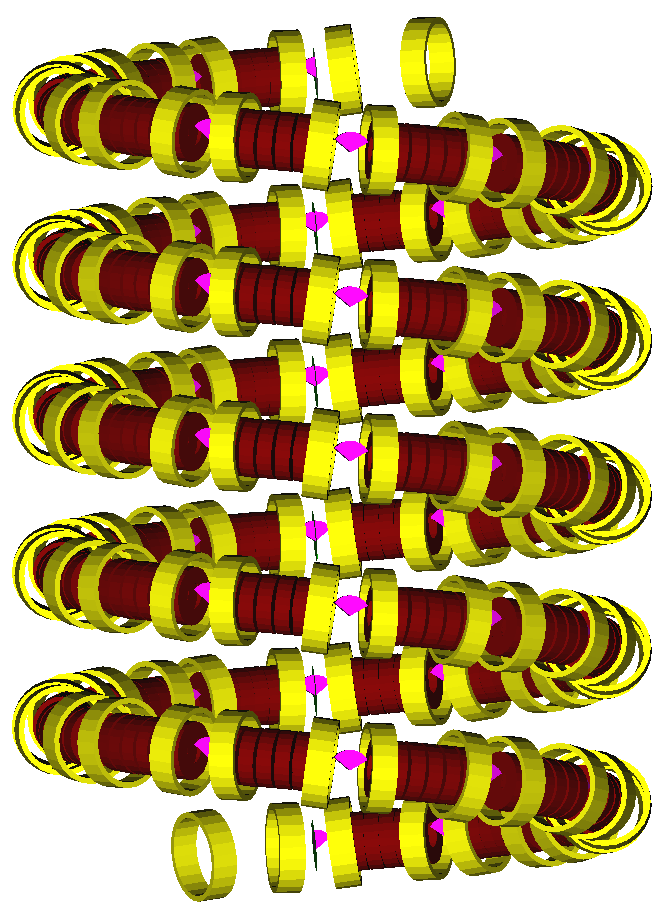}
	\caption{Multilayer Guggenheim channel layout. Color-coding is the same as for Fig.~\ref{fig:rfofo}.}
	\label{fig:guggenheim}
\end{figure}

Figure~\ref{fig:guggenheim} shows the 5-turn layout which has been simulated. Along with the unshielded case with all the magnetic coils of all layers contributing to the magnetic field guiding the muons, another scheme has been considered, with shielding between the individual layers. Both layouts include safety windows around absorbers and beryllium windows in the RF cavities. 

\begin{figure}
	\centering
		\includegraphics[width=0.48\textwidth]{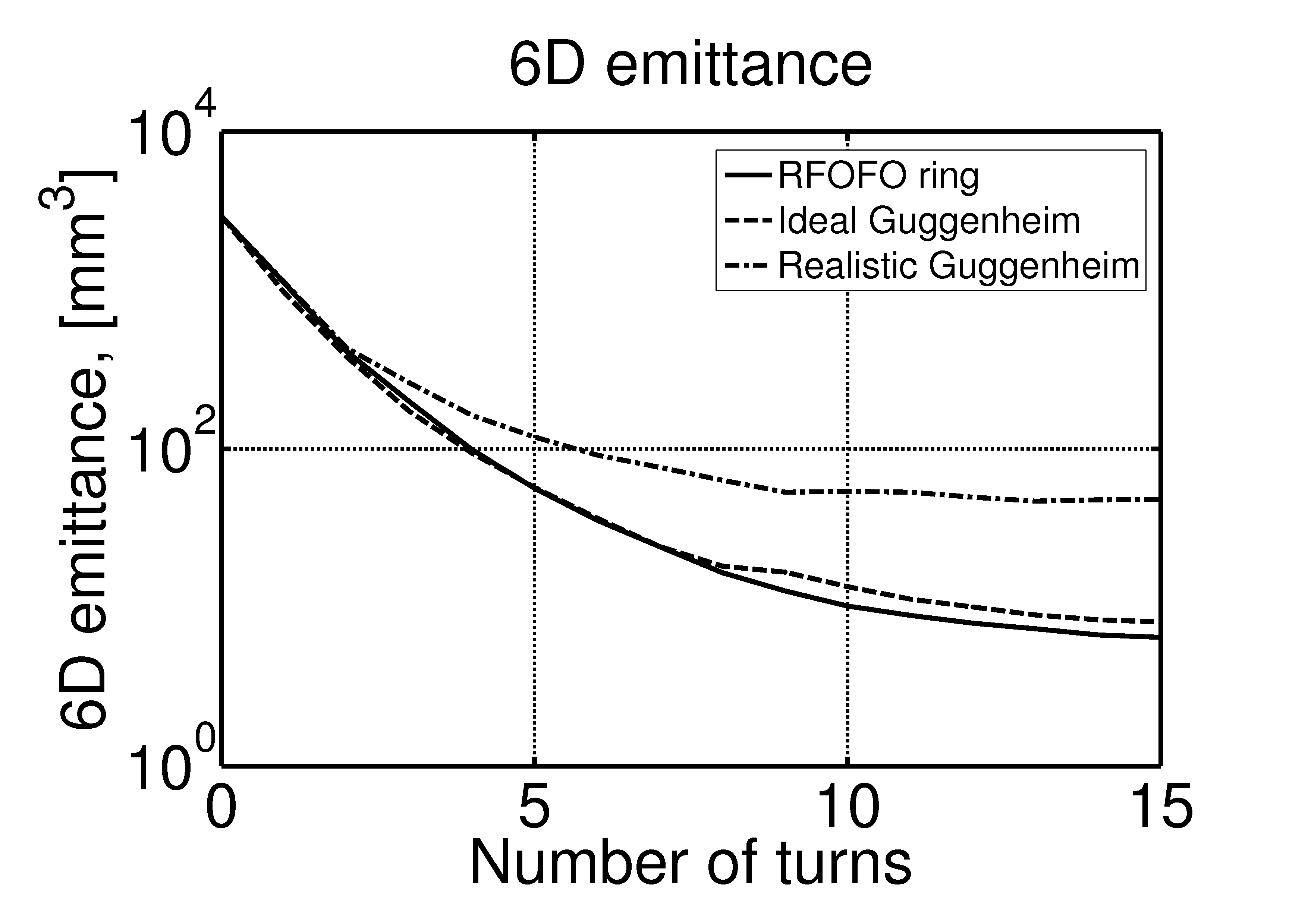}
	\caption{Six-dimensional emittance reduction \emph{vs.} number of turns.}
	\label{fig:6d}
\end{figure}

\begin{figure}
	\centering
		\includegraphics[width=0.48\textwidth]{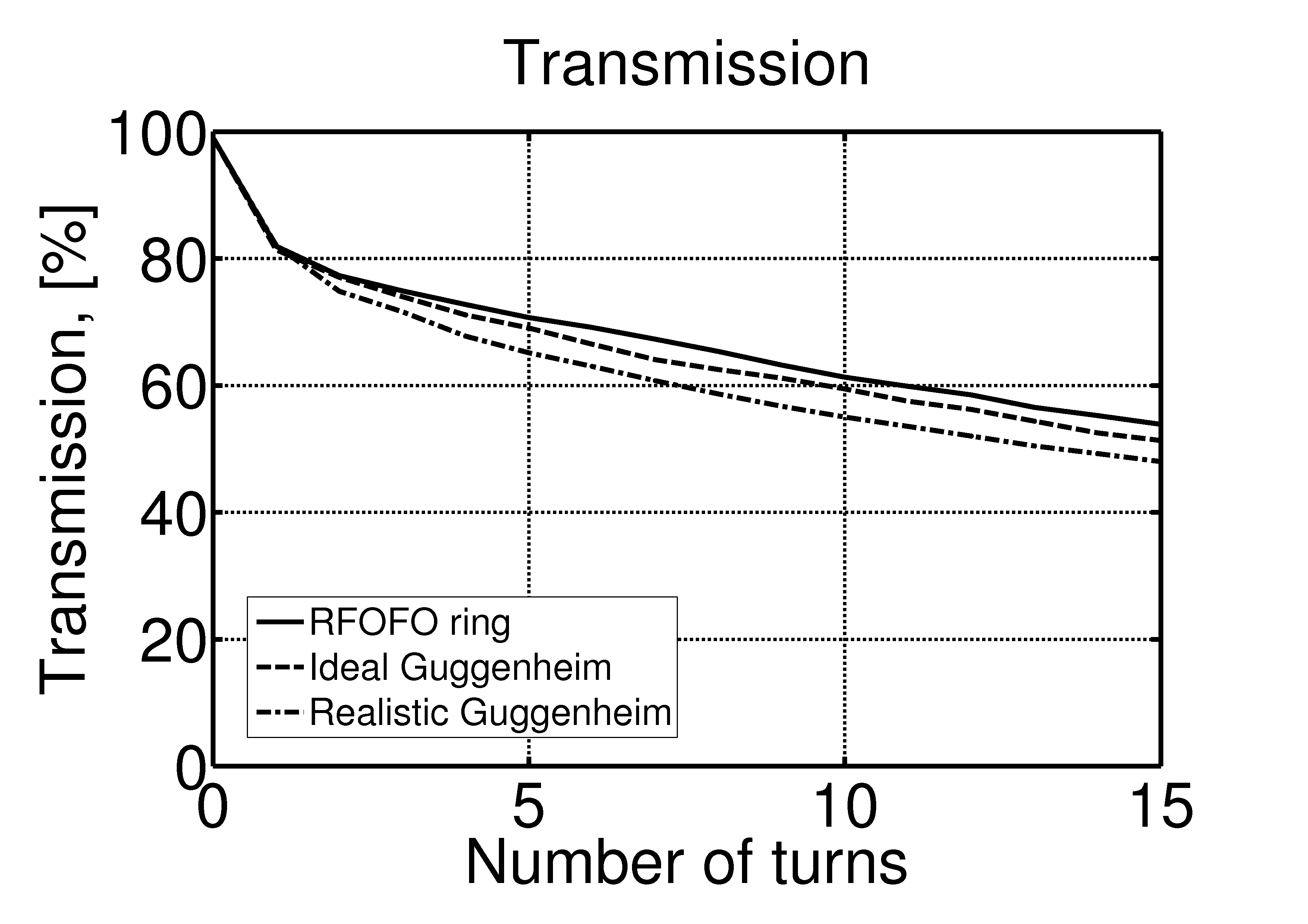}
	\caption{Transmission---percent of surviving muons \emph{vs.} number of turns.}
	\label{fig:transmission}
\end{figure}

The simulation details can be found in \cite{Guggenheim}. Here we show only the six-dimensional emittance reduction (see Fig.~\ref{fig:6d}) and the transmission (see Fig.~\ref{fig:transmission}) as functions of the number of turns. The transmission is measured as the ratio of the number of particles at a certain arclength to the initial number of particles. Muon decay and stochastic processes are taken into account. The solid line is used for the RFOFO ring, which serves as a reference, the dashed line represents the Guggenheim channel with shielding between layers and no windows in absorbers or RF cavities (the idealized Guggenheim, the performance of which should not differ significantly from the RFOFO ring, which is indeed the case), and the dash-dotted line represents the realistic Guggenheim with shielding between layers and windows in both absorbers and RF cavities. 

Figure~\ref{fig:6d} clearly demonstrates significant six-dimensional cooling; however, the performance of the cooling channel is seriously affected by the use of the absorber and RF windows. These results are in agreement with earlier studies for the RFOFO ring \cite{PhysRevSTAB.8.061003}.

\section{RF Breakdown and the Open Cavity Lattice}

\begin{figure}
	\centering
		\includegraphics[width=0.48\textwidth]{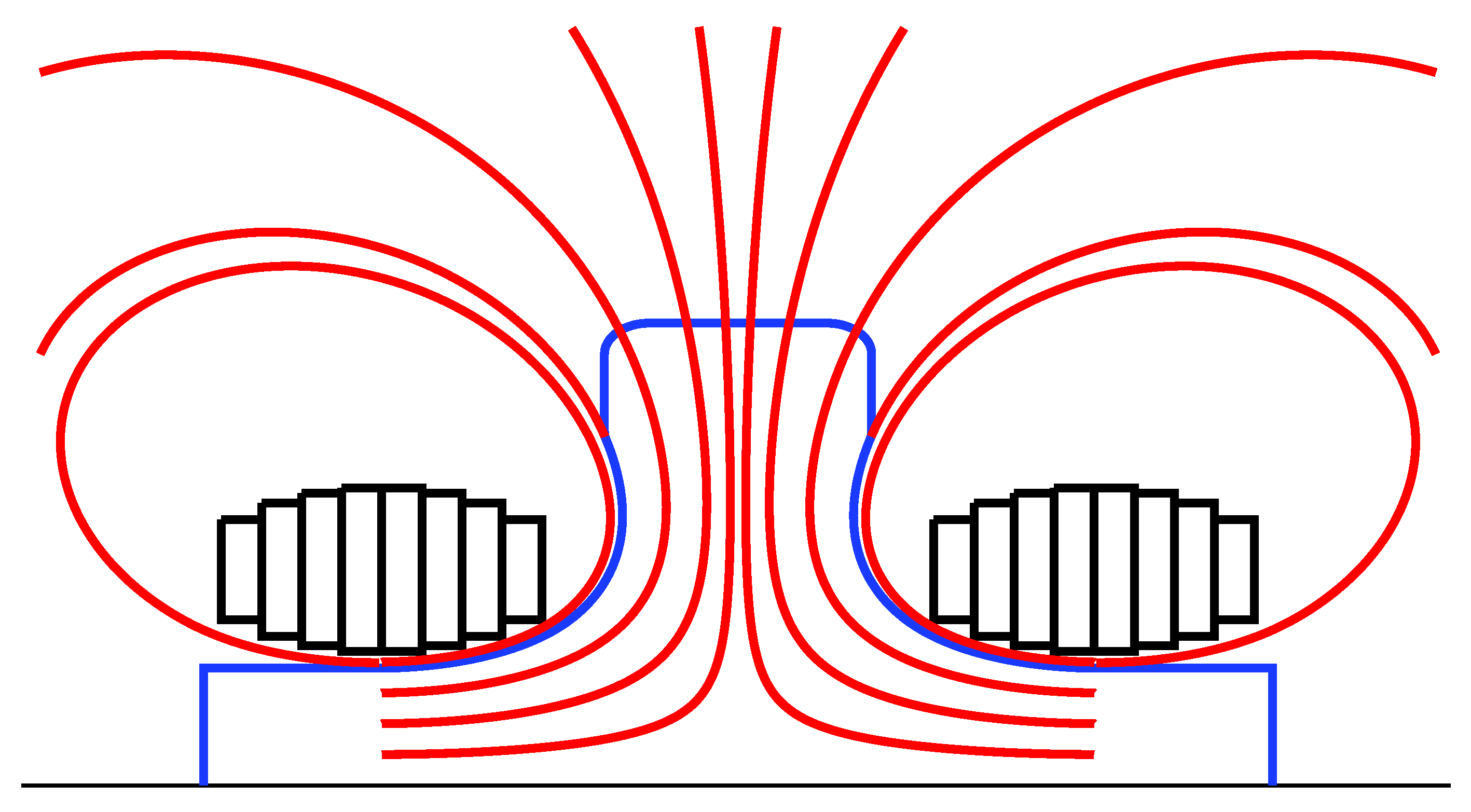}
	\caption{Open cavity shape. Black---coils generating the magnetic field, blue---walls of the RF cavity, red---magnetic field lines.}
	\label{fig:opencavityB}
\end{figure}

\begin{figure}
	\centering
		\includegraphics[width=0.48\textwidth]{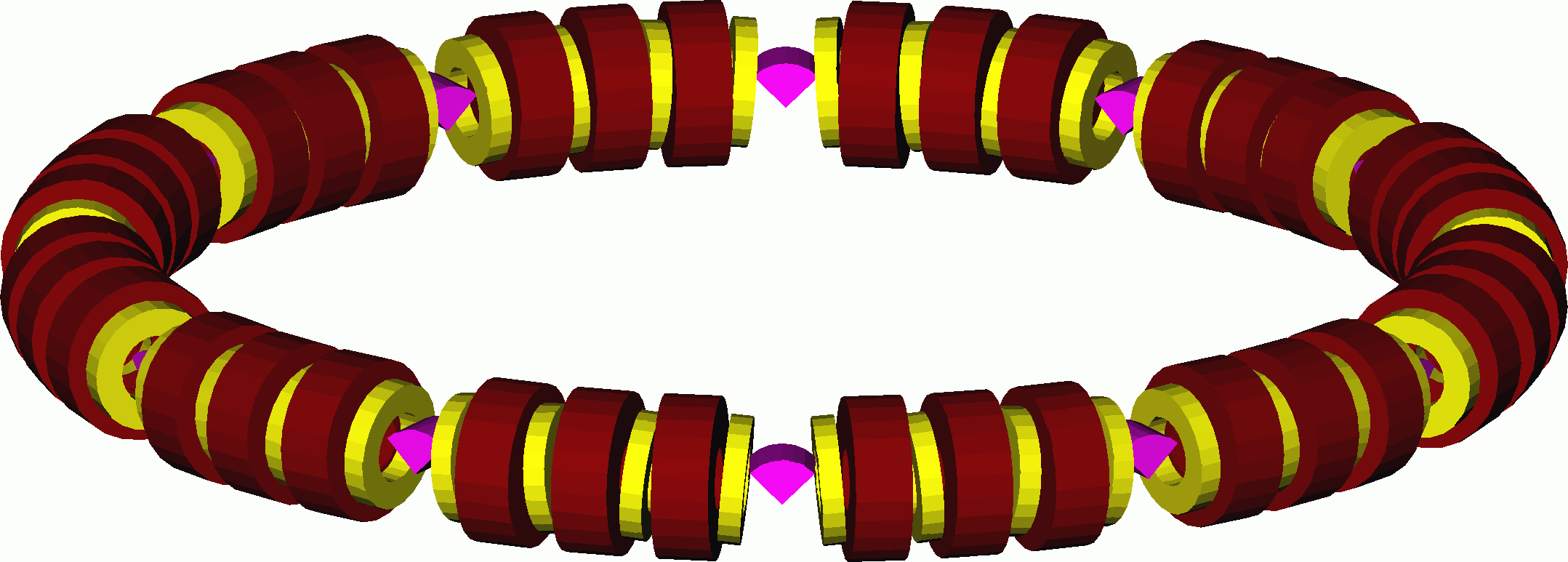}
	\caption{Open cavity layout. Yellow---tilted magnetic coils to provide necessary bending and focusing and generate dispersion, magenta---wedge absorbers for cooling and emittance exchange, red---RF cavities for restoring the longitudinal component of the momentum.}
	\label{fig:open_cavity}
\end{figure}

Various studies suggest that the presence of the magnetic field disrupts the performance of RF cavities by causing breakdown \cite{norem,rf_breakdown}. Thus, it was proposed to consider an alternative layout of the cooling channel, the so-called open cavity lattice \cite{open_cavity}. The concept itself consists of two parts: a) moving the solenoidal coils from over the RF cavities into the irises; and b) shaping the RF cavities such that the walls of the cavities are predominantly parallel to the magnetic field lines (see Fig.~\ref{fig:opencavityB}), which hopefully solves the problem of the breakdown. The current layout illustrated in Figure~\ref{fig:open_cavity} does not include specifically shaped RF cavities; instead a simplified pillbox geometry is used. 

The main parameters of the open cavity lattice are summarized in Table~\ref{tab:params_open_cavity} and compared to the parameters of the original RFOFO channel.

\begin{table}
	\centering
\begin{footnotesize}
		\begin{tabular}{c|c|c|c}
			& & Open cavity	& RFOFO\\
		  \hline\hline
			RF frequency & [MHz] & 201.25	& 201.25\\
			RF gradient & [MV/m] & 16.075 & 12.835\\
			Maximum axial field & [T]	& 3.23 & 2.80\\
			Circumference & [m]	& 30.72	& 33.00\\
			Radius & [m]	& 4.889 & 5.252\\
			Reference momentum & [MeV/c] & 214 & 201\\
			Coil tilt & [deg] & 4.90 & 3.04\\
			Current densities & [A/mm$^2$] & $[63,45,-45,-63]$ & $[95,-95]$ \\
			Absorber angle & [deg] & 90 & 110\\
			Absorber vertical offset & [cm] & 12.0 & 9.5\\
			Absorber axial length & [cm] & 24.00 & 27.13\\
		\end{tabular}
\end{footnotesize}
	\caption{Parameters of the open cavity lattice compared to the RFOFO ring lattice.}
	\label{tab:params_open_cavity}
\end{table}

The new cooling ring has 12 cells with three RF cavities in each and four solenoidal coils in the irises. These coils bear currents with the following densities: 63 A/mm$^2$, 45 A/mm$^2$, $-45$ A/mm$^2$, $-63$ A/mm$^2$. 

The circumference of the ring is 30.72 meters. The idea of tipping the solenoids, similar to the RFOFO ring concept, is employed in this layout to generate an average vertical magnetic field of 0.136 T providing necessary bending. Solenoid axes are tilted $4.9^\circ$ above or below the orbital midplane depending on the direction of the current. The centers of the solenoids are displaced radially outward from the reference circle by 21 mm to minimize the integrated on-axis radial field, thus, vertical beam deviations. This technique allows to save 2\% of the beam that would be lost with no offset. 

The fact that the solenoids are tilted leads to the reduction of the amount of space available for the RF system; hence, the energy gain per cell is limited, which, in turn, limits the angle of the wedge absorber to approximately $90^\circ$ (compared to $110^\circ$ in RFOFO).

\begin{figure}
	\centering
		\includegraphics[width=0.48\textwidth]{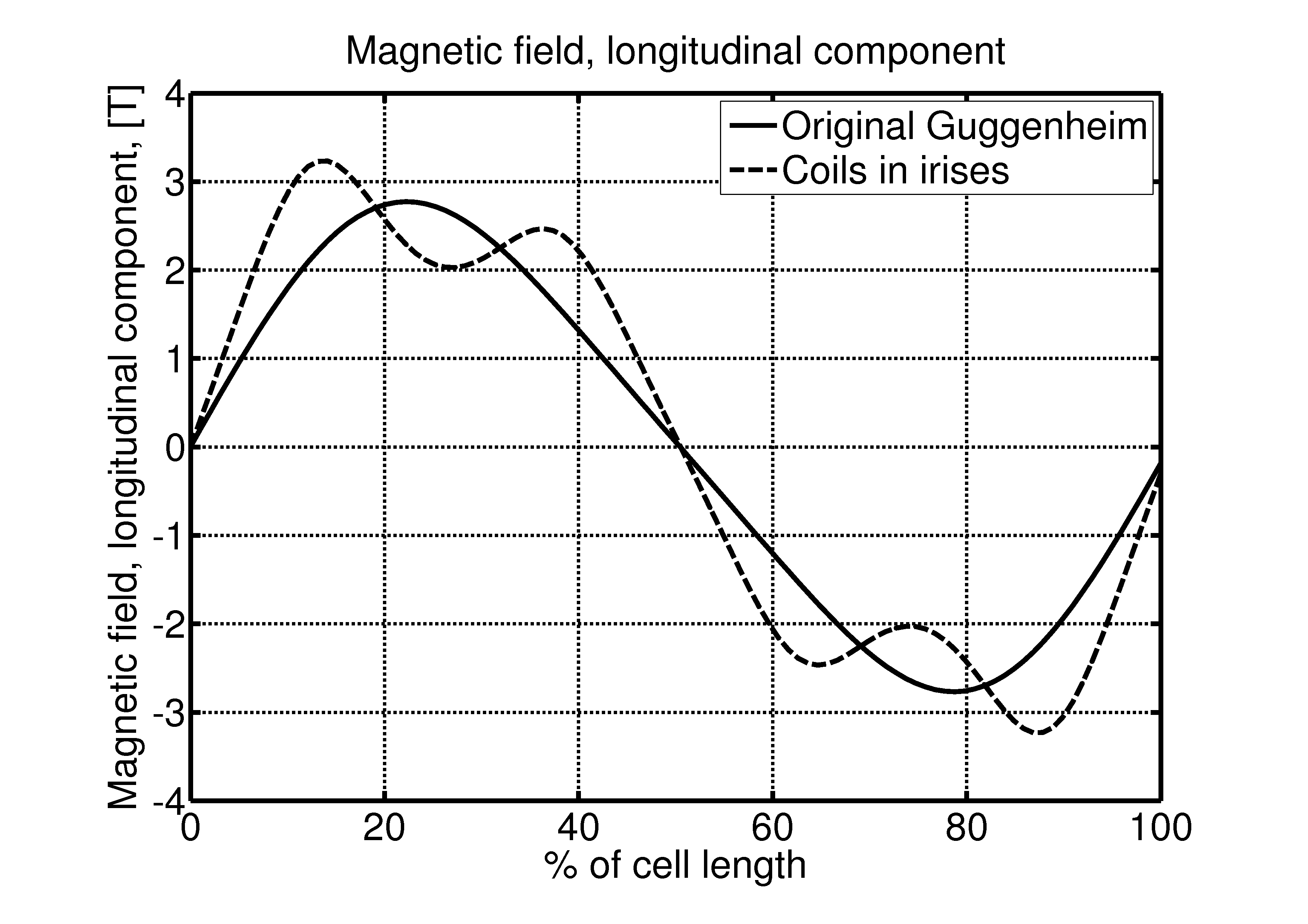}
	\caption{Longitudinal field component. Solid line---original RFOFO ring or Guggenheim helix lattice, dashed line---open cavity lattice.}
	\label{fig:long}
\end{figure}

\begin{figure}
	\centering
		\includegraphics[width=0.48\textwidth]{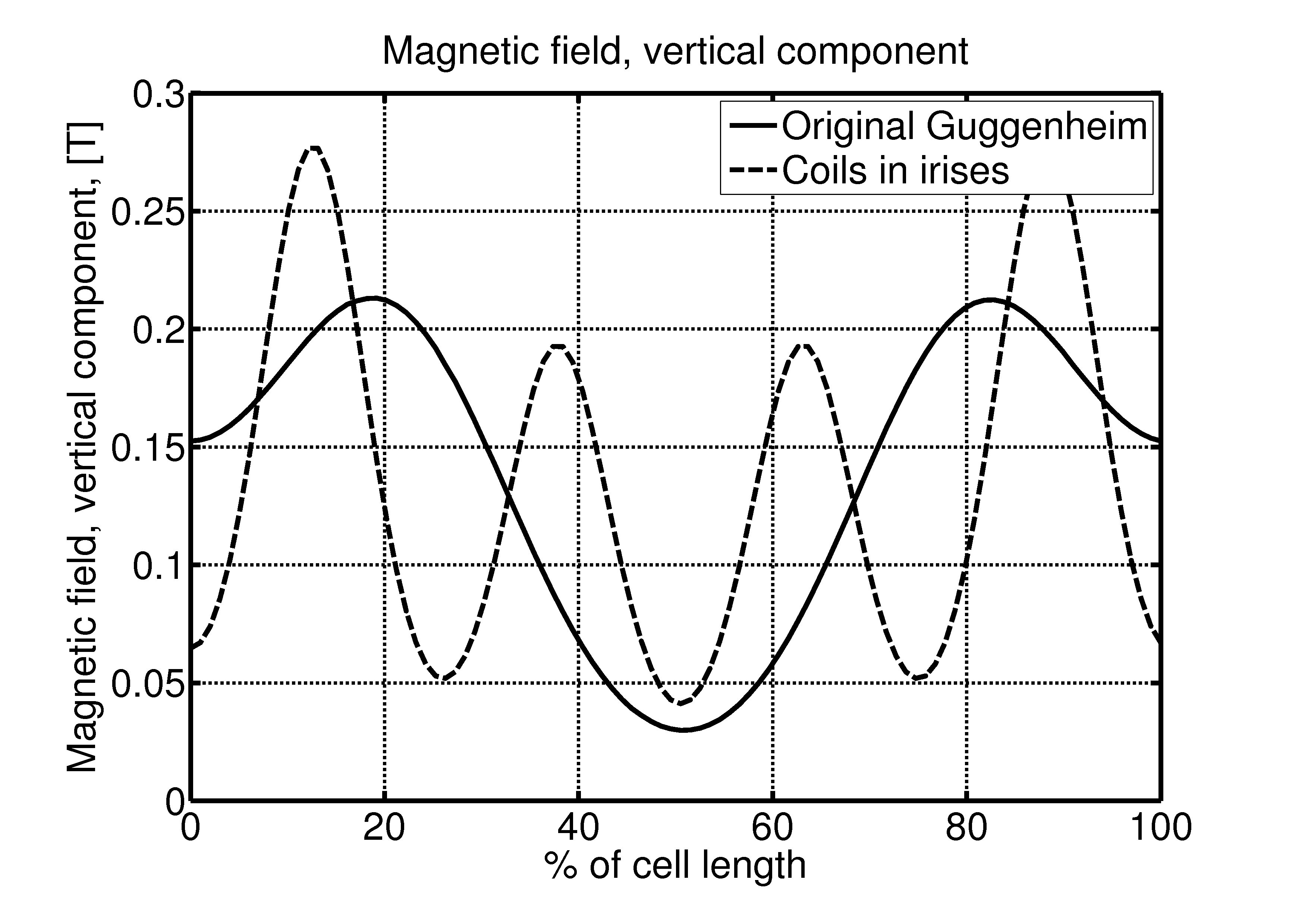}
	\caption{Vertical field component. Solid line---original RFOFO ring or Guggenheim helix lattice, dashed line---open cavity lattice.}
	\label{fig:vert}
\end{figure}

\begin{figure}
	\centering
		\includegraphics[width=0.48\textwidth]{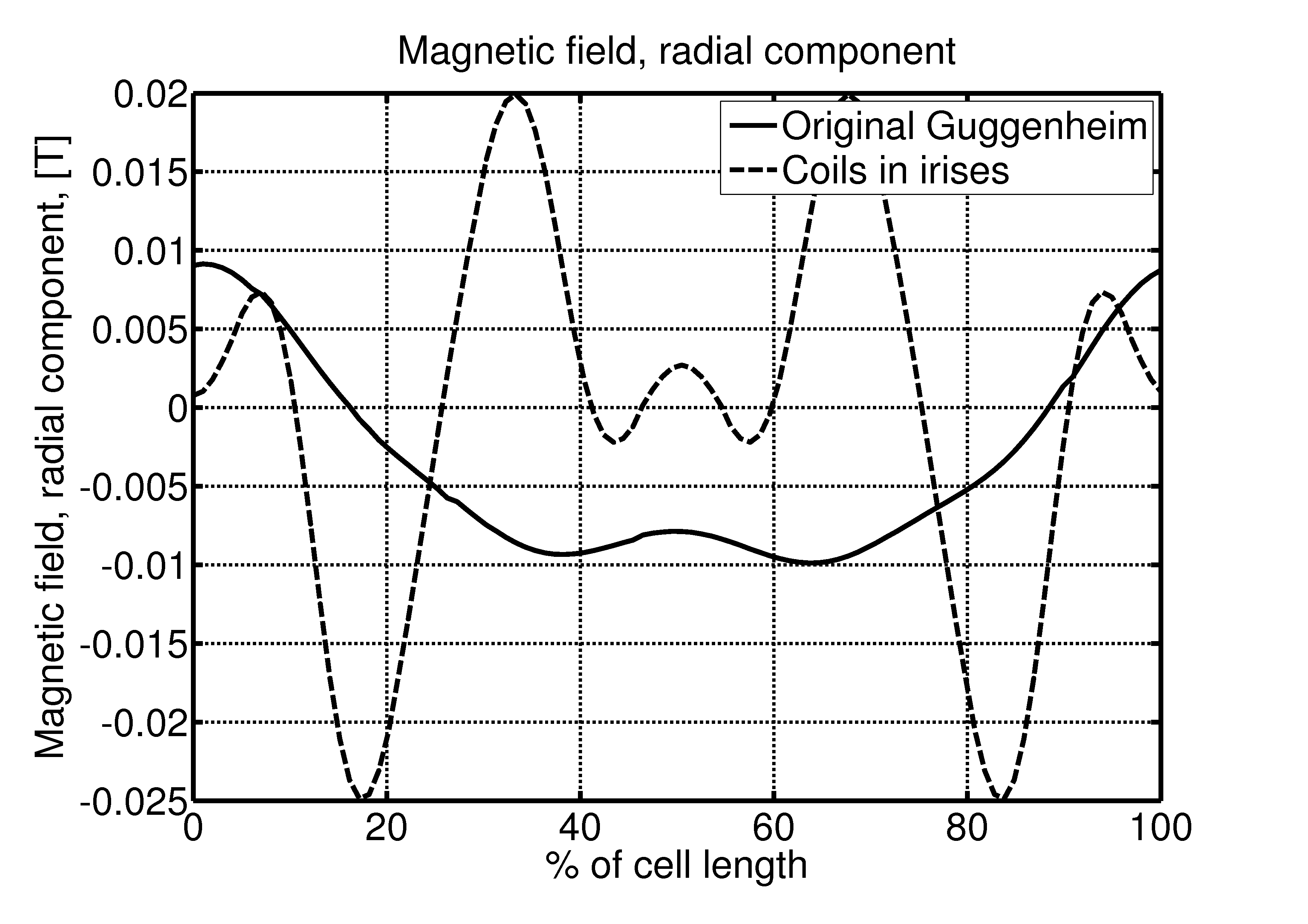}
	\caption{Radial field component. Solid line---original RFOFO ring or Guggenheim helix lattice, dashed line---open cavity lattice.}
	\label{fig:rad}
\end{figure}

Figures~\ref{fig:long}--\ref{fig:rad} illustrate the difference between field components for the original RFOFO design and the new design with coils in irises. Since there are four coils per cell, all field profiles have more complicated shapes; however, the overall magnitudes are similar. 

The peak in the longitudinal field is still approximately 3 T; the radial component is more pronounced, but still small compared to both the vertical and the longitudinal components. The vertical component is everywhere positive providing an average bending field of 0.136 T. 

\begin{figure}
	\centering
		\includegraphics[width=0.48\textwidth]{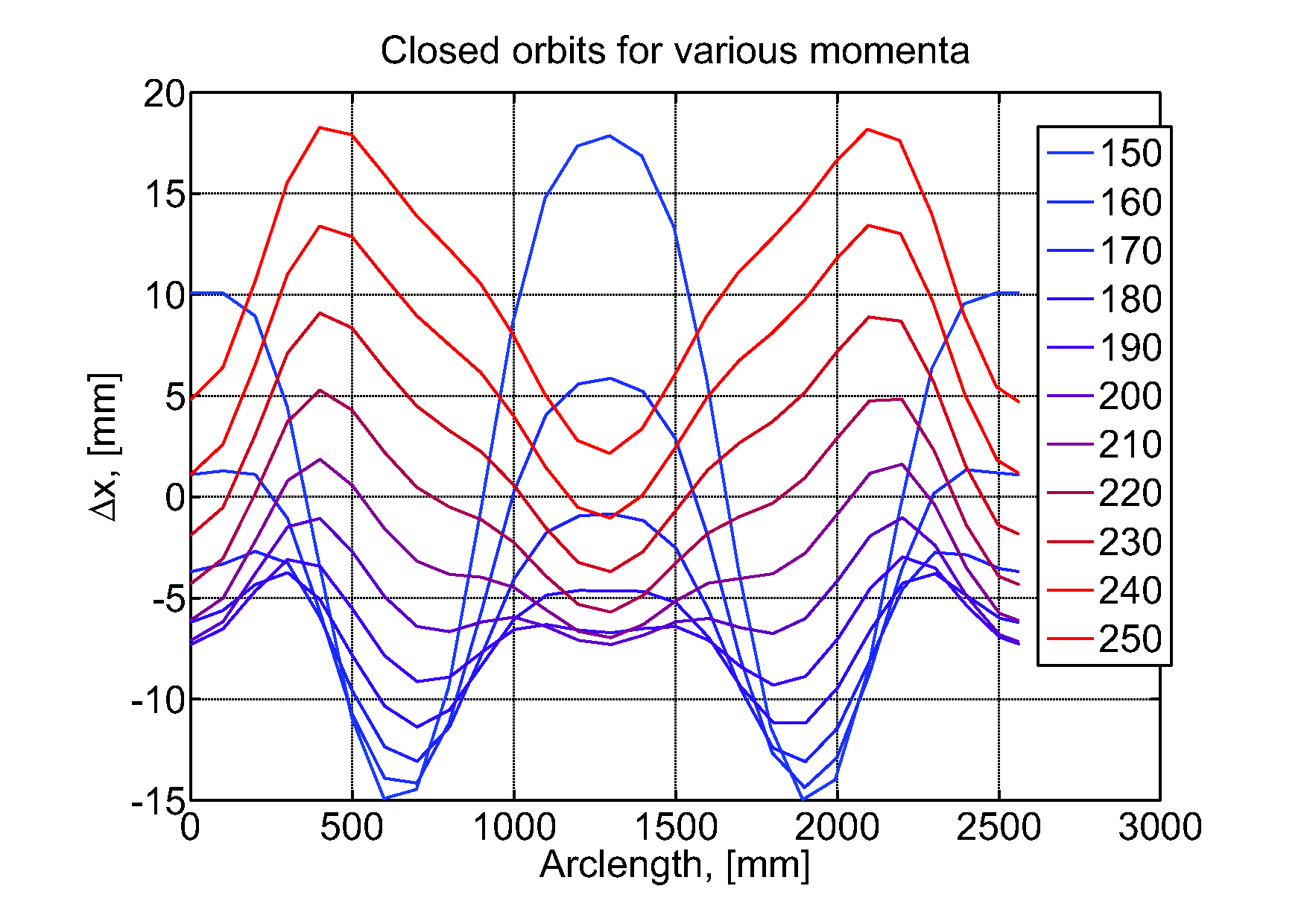}
	\caption{Closed orbit horizontal offset along one cell of the cooling channel (2560 mm) for various momenta from 150 MeV/c to 250 MeV/c.}
	\label{fig:closed_orbit_x}
\end{figure}

\begin{figure}
	\centering
		\includegraphics[width=0.48\textwidth]{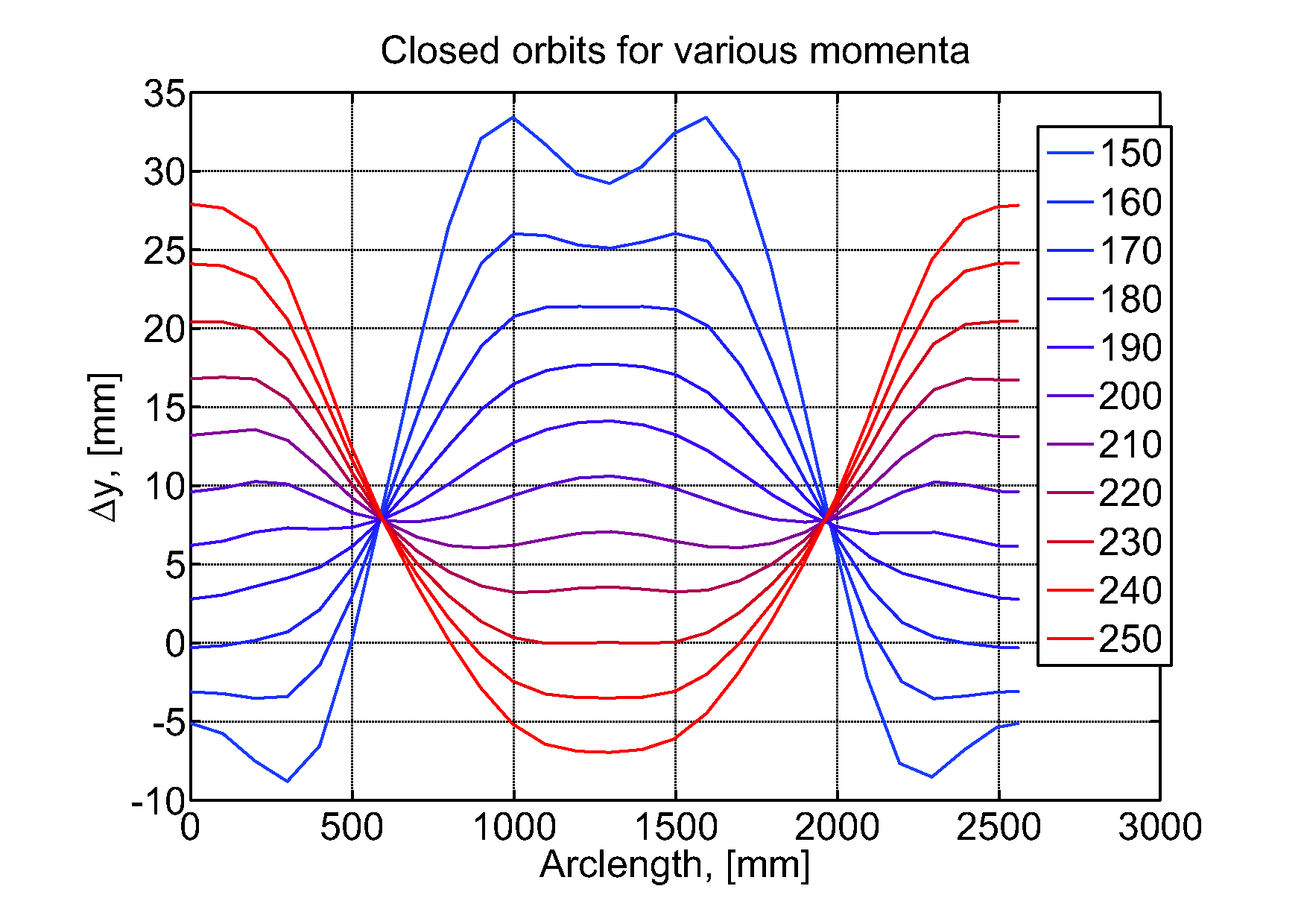}
	\caption{Closed orbit vertical offset along one cell of the cooling channel (2560 mm) for various momenta from 150 MeV/c to 250 MeV/c.}
	\label{fig:closed_orbit_y}
\end{figure}

Deviation of the closed orbit along one periodic cell for various momenta ranging from 150 MeV/c to 250 MeV/c are shown in Figs.~\ref{fig:closed_orbit_x}--\ref{fig:closed_orbit_y}. 

\begin{figure}
	\centering
		\includegraphics[width=0.48\textwidth]{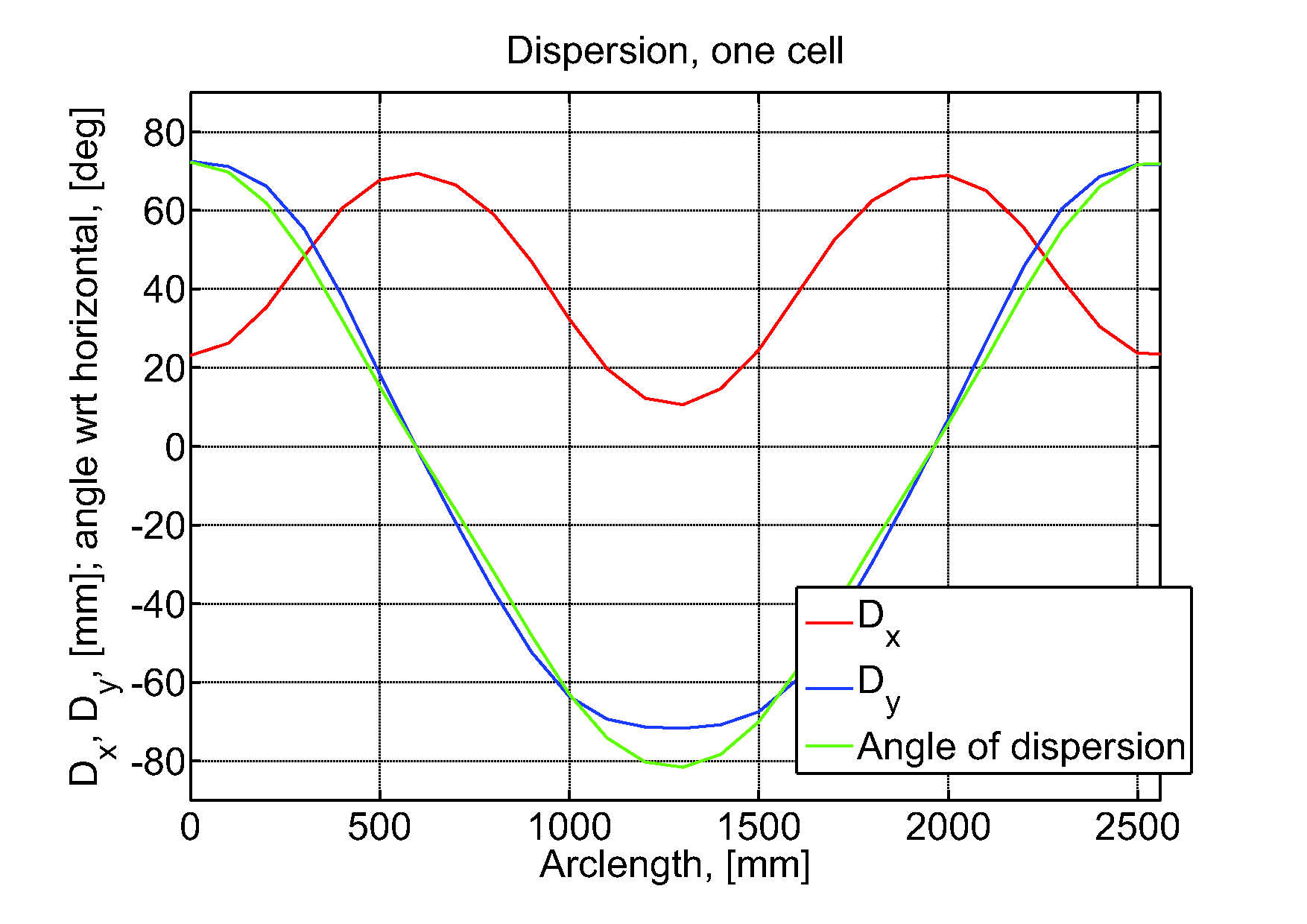}
	\caption{Dispersion plot. Red line---horizontal, blue line---vertical, green---orientation of dispersion with respect to the horizontal axis.}
	\label{fig:dispersion}
\end{figure}

The dispersion calculation (Fig.~\ref{fig:dispersion}) shows that the dispersion at the absorber plane (beginning of the cell) is primarily in the vertical direction, at an angle of $\sim\!\!20^\circ$ from the vertical axis. This fact suggests the orientation of the absorbers. The dispersion in the the center of the cell is negative, again mainly in the vertical direction.

\subsection{Dynamics of particles in the open cavity lattice}

\begin{figure}
	\centering
		\includegraphics[width=0.48\textwidth]{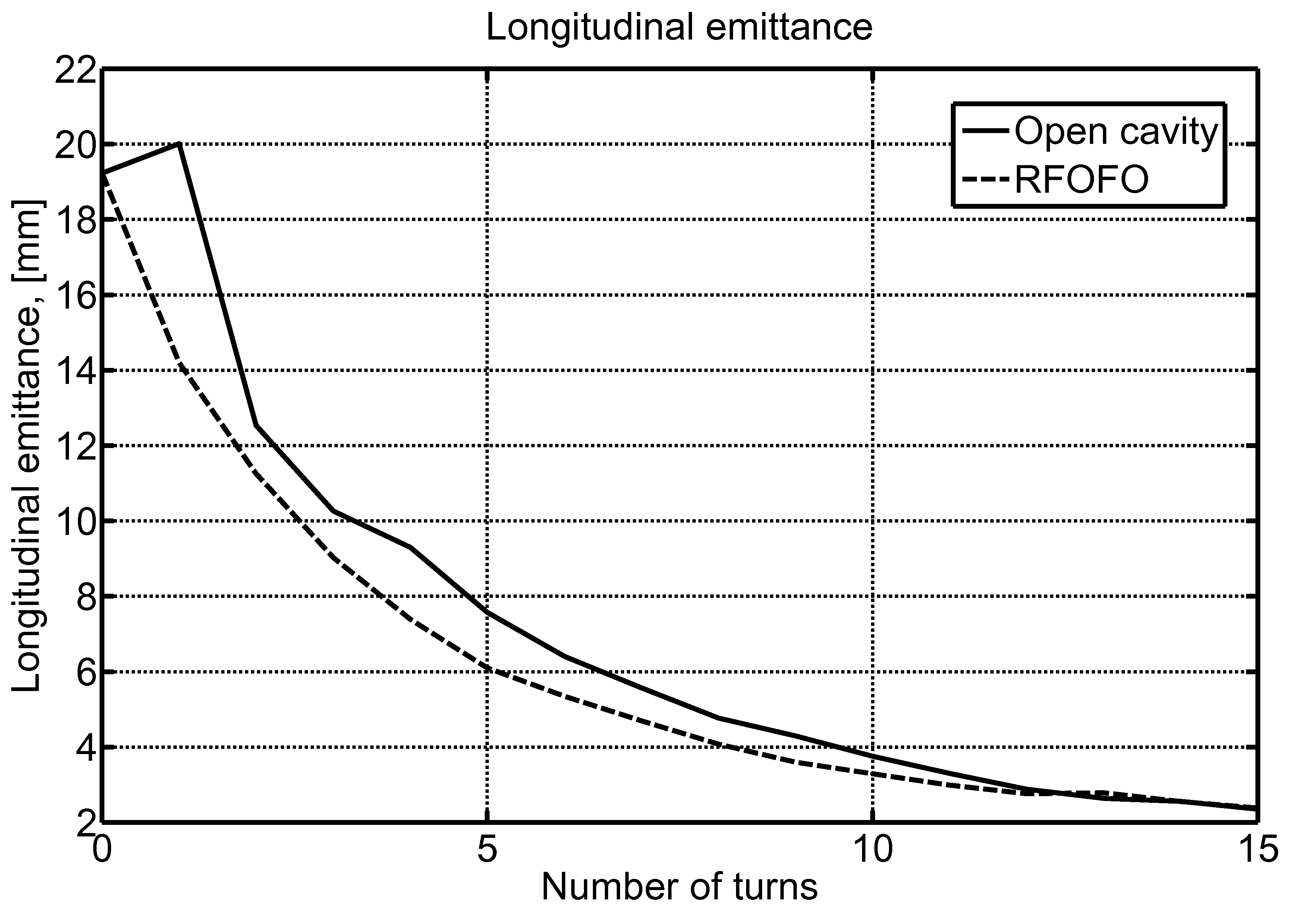}
	\caption{Longitudinal emittance reduction vs number of turns. Solid line---open cavity lattice, dashed line---RFOFO.}
	\label{fig:long_decay}
\end{figure}

\begin{figure}
	\centering
		\includegraphics[width=0.48\textwidth]{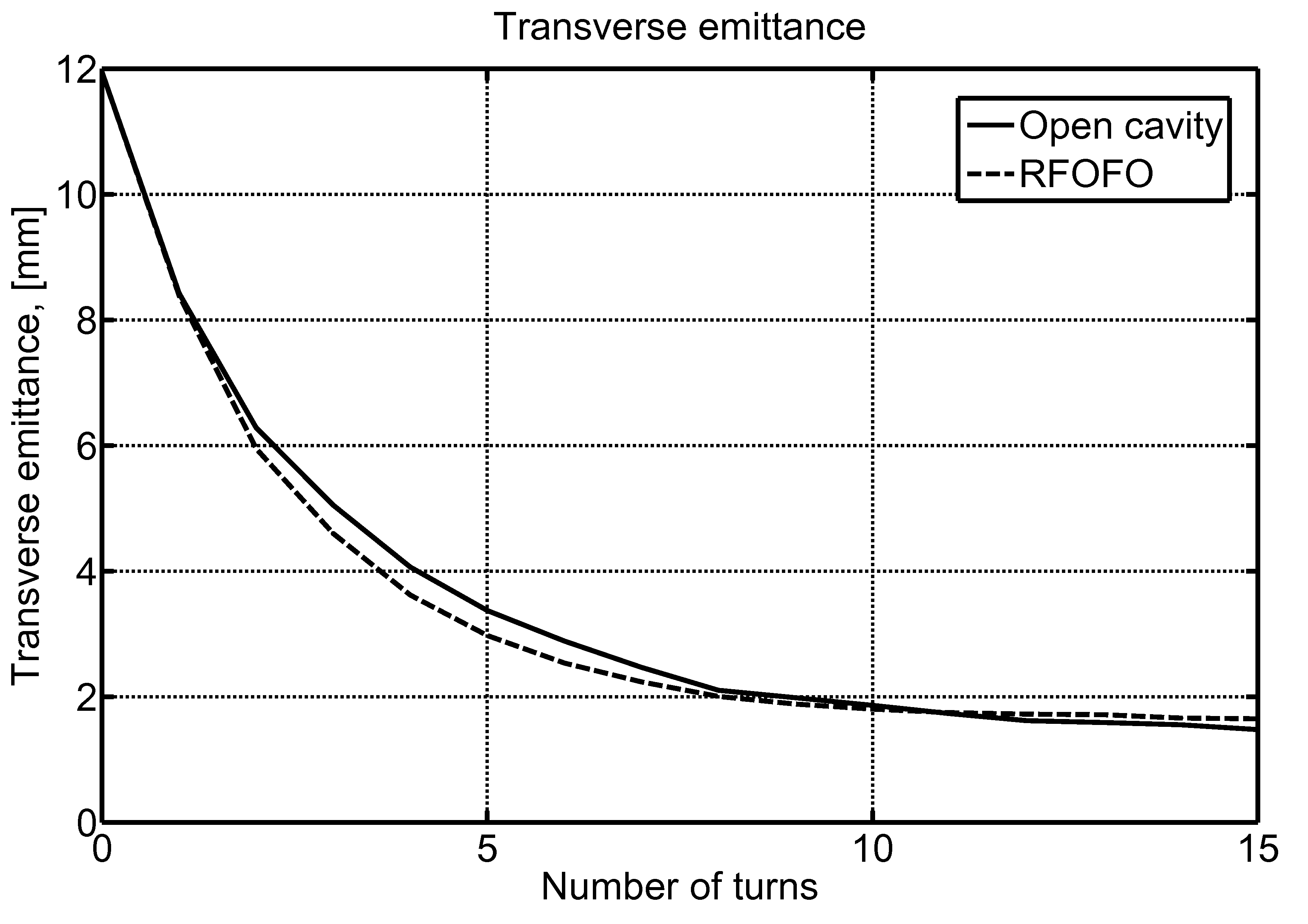}
	\caption{Transverse emittance reduction vs number of turns. Solid line---open cavity lattice, dashed line---RFOFO.}
	\label{fig:trans_decay}
\end{figure}

\begin{figure}
	\centering
		\includegraphics[width=0.48\textwidth]{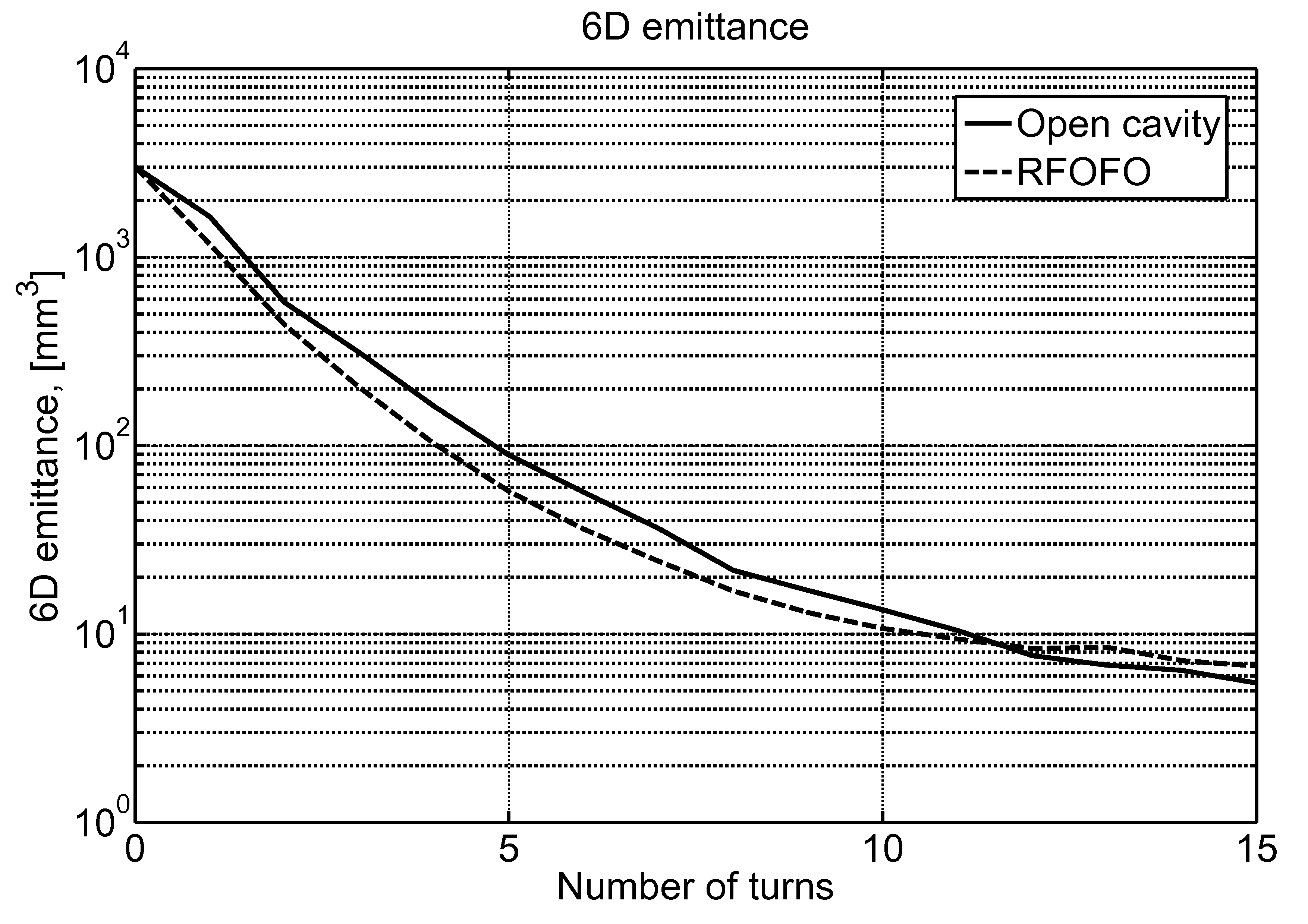}
	\caption{Six-dimensional emittance reduction vs number of turns. Solid line---open cavity lattice, dashed line---RFOFO.}
	\label{fig:6d_decay}
\end{figure}

\begin{figure}
	\centering
		\includegraphics[width=0.48\textwidth]{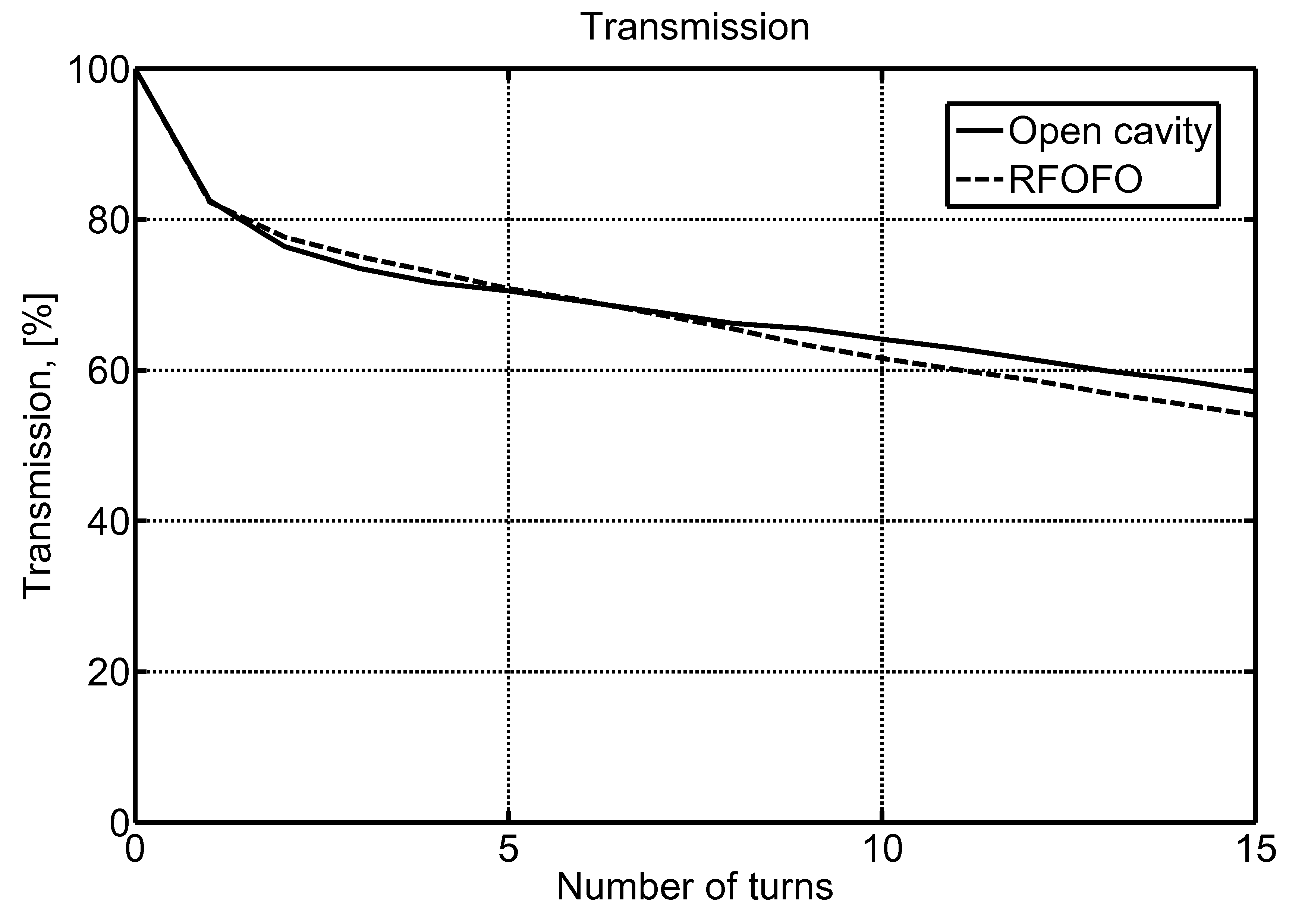}
	\caption{Transmission---percent of surviving muons vs number of turns. Solid line---open cavity lattice, dashed line---RFOFO.}
	\label{fig:transmission_decay}
\end{figure}

In this section we compare the performance of the open cavity lattice with that of the RFOFO ring discussed in Section 1. Figures \ref{fig:long_decay}--\ref{fig:transmission_decay} illustrate the change in the longitudinal, transverse and six-dimensional emittance versus the number of turns for both the open cavity lattice and the RFOFO ring with all the decay and stochastic processes taken into account. As one can see from these tracking results, the open cavity lattice clearly performs very similarly to the RFOFO lattice, sometimes even slightly better for the same initial beam. The open cavity ring is slightly smaller than the RFOFO ring: 30.72 m in circumference vs 33.00 m, respectively. However, there is less room for RF cavities in the open cavity layout, hence, higher gradients are required (as can be seen from Table~\ref{tab:params_open_cavity}).

\begin{figure}
	\centering
		\includegraphics[width=0.48\textwidth]{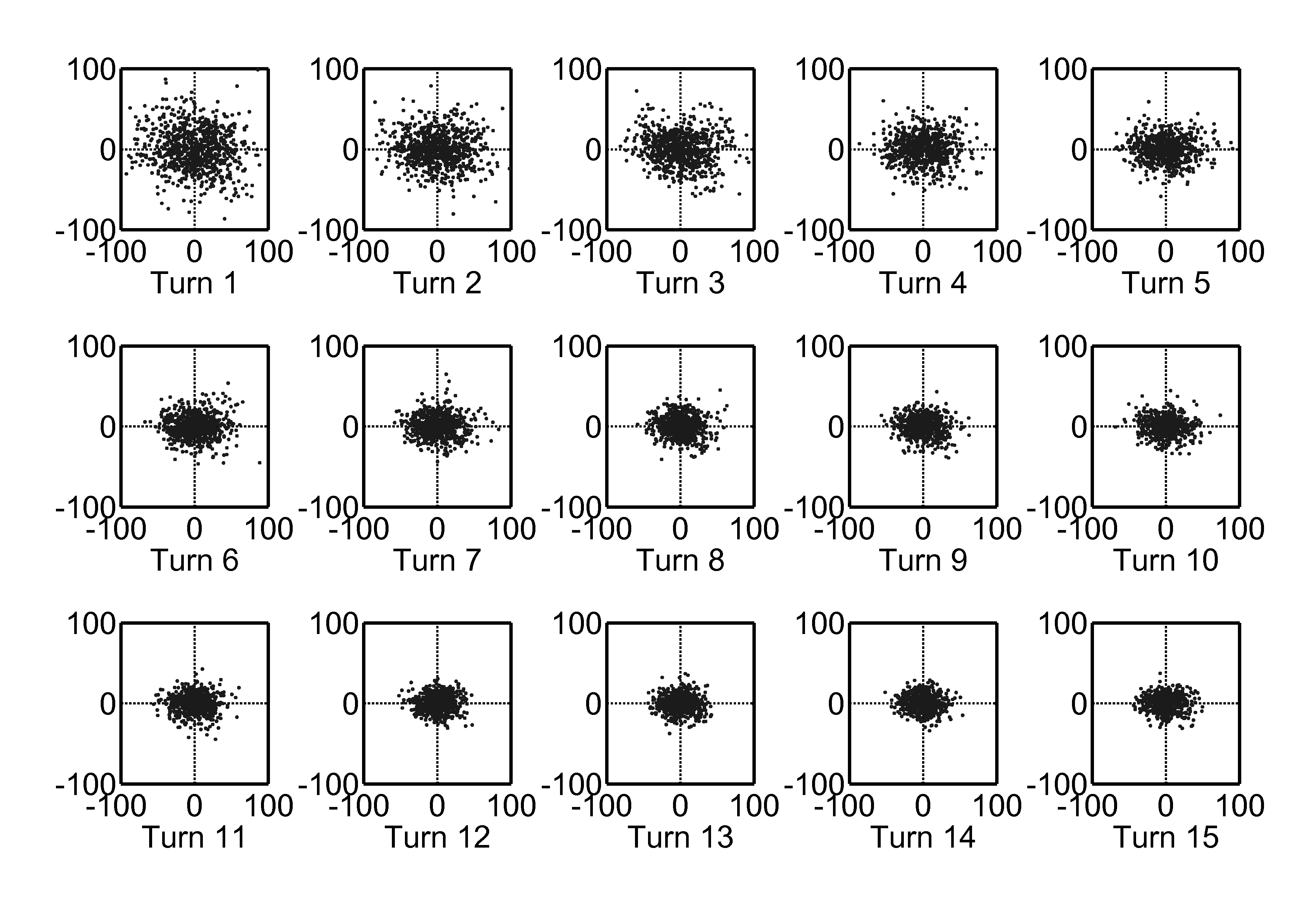}
	\caption{Phase portraits in the $(x-p_x)$ plane for different number of turns.}
	\label{fig:cooling_xpx_decay}
\end{figure}

\begin{figure}
	\centering
		\includegraphics[width=0.48\textwidth]{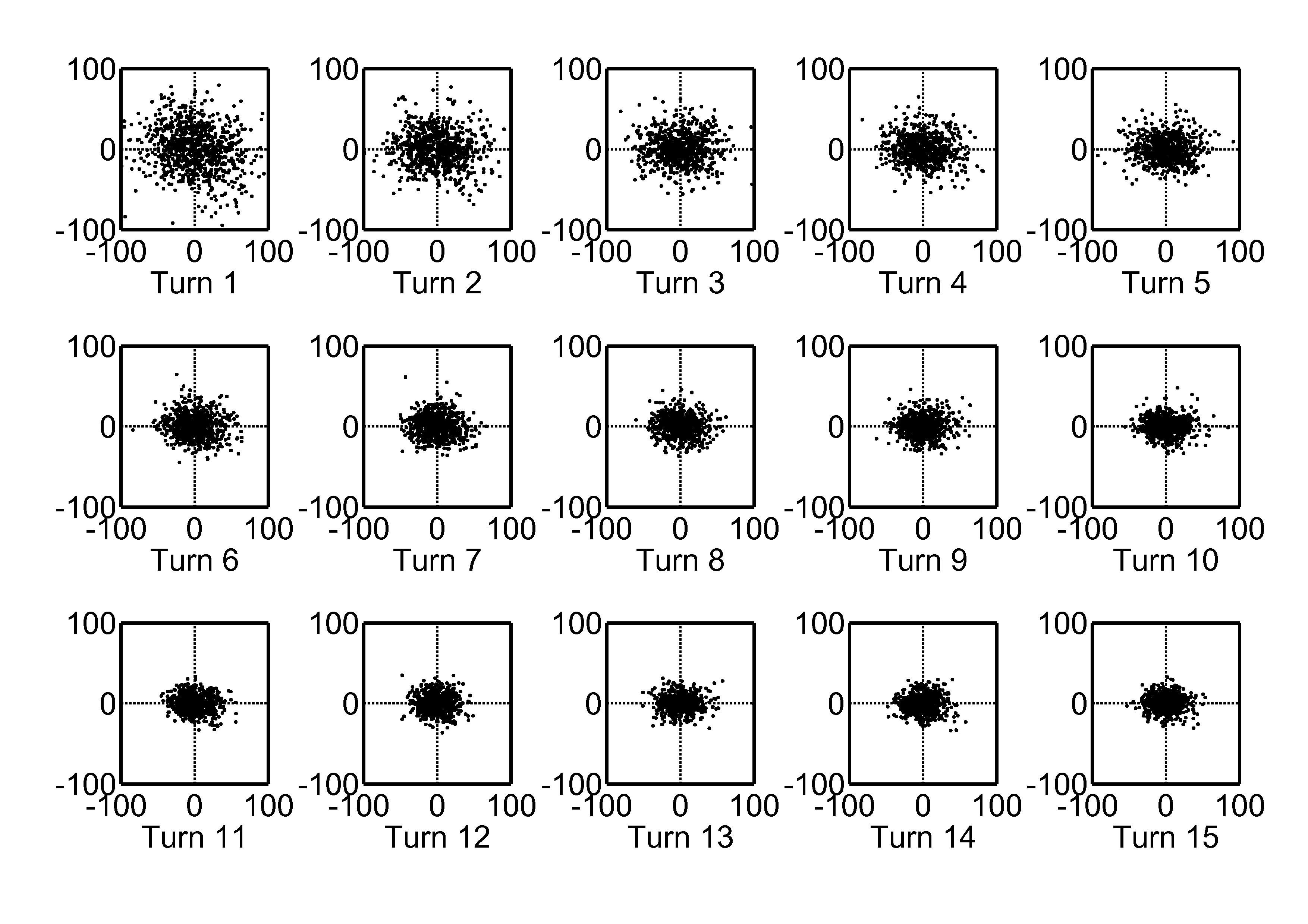}
	\caption{Phase portraits in the $(y-p_y)$ plane for different number of turns.}
	\label{fig:cooling_ypy_decay}
\end{figure}

\begin{figure}
	\centering
		\includegraphics[width=0.48\textwidth]{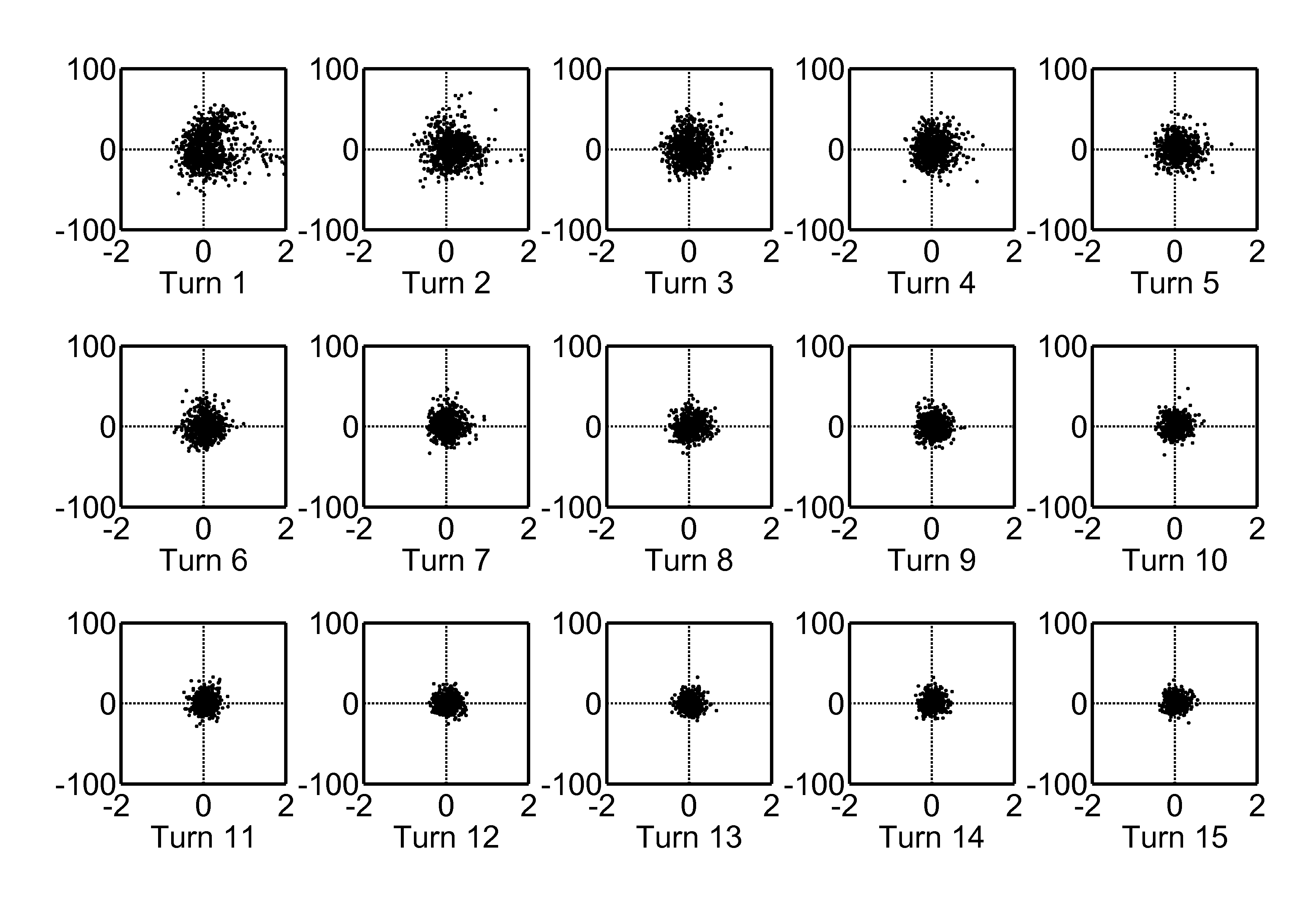}
	\caption{Phase portraits in the $(t-p_z)$ plane for different number of turns.}
	\label{fig:cooling_tpz_decay}
\end{figure}

The cooling effect in both transverse and longitudinal directions is evident from Figs.~\ref{fig:cooling_xpx_decay}--\ref{fig:cooling_tpz_decay}.

An alternative approach to using the open cavity or any other type of magnetically insulating lattice is to research the techniques allowing RF cavities to withstand higher magnetic field without breaking down. Such techniques include atomic layer deposition creating a thin layer of material on the cavity walls \cite{norem_pac2009}, high-pressure gas filled cavities \cite{chung}, dielectric-loaded cavities \cite{popovic_pac2009}, and using other materials such as aluminum or beryllium and low temperatures \cite{palmer_materials}.

\begin{acknowledgments}
This work is supported by the United States Department of Energy under grant No.~\mbox{DE-FG02-07ER41487}.

The authors would like to thank Richard Fernow, Robert Palmer and Akira Sato for fruitful and inspiring discussions and new ideas.
\end{acknowledgments}


\end{document}